\documentclass[10pt,showpacs,preprintnumbers,amsmath,amssymb,
aps,prd,nofootinbib,eqsecnum,a4paper]{revtex4}
\usepackage{graphicx,epsf,bm}
\usepackage{amsmath, subfigure}  
\def\be{\begin{equation}}
\def\ee{\end{equation}}
\def\beb{\begin{equation*}}
\def\eeb{\end{equation*}}
\def\bea{\begin{eqnarray}}
\def\eea{\end{eqnarray}}
\def\beab{\begin{eqnarray*}}
\def\eeab{\end{eqnarray*}}
\def\nn{\nonumber}

\def\bi{\begin{itemize}}
\def\ei{\end{itemize}}

\def\w{{\omega}}
\def\cs2{c_{\rm{s}}^2}

\def \beg {\begin{enumerate}}
\def \en {\end{enumerate}}

\def\Pb{P_0}
\def\rhob{\rho_0}

\def\drho{{\delta\rho}}
\def\dP{{\delta P}}
\def\dPn{{\delta P_{\rm{nad}}}}
\def\dPnad{{\delta P_{1\rm{nad}}}}

\def\cs{c_{\rm{s}}^2}

\def\p{\partial}

\def\H{{\cal H}}
\def\S{{\cal{S}}}

\newcommand\eq[1]{Eq.~(\ref{#1})}

\newcommand\Ref[1]{Ref.~\cite{#1}}
\def\U0{{\bar U_0}}
\def\V1{{\bar{V_1}}}

\begin{document}
\preprint{} 
\title{Estimating the amount of vorticity generated by cosmological
perturbations in the early universe}
\author{Adam J.~Christopherson$^1$}
\email{a.christopherson@qmul.ac.uk}
\author{Karim A.~Malik$^1$}
\author{David R.~Matravers$^2$}
\affiliation{
$^1$Astronomy Unit, School of Mathematical Sciences, Queen Mary University
of London, Mile End Road, London, E1 4NS, United Kingdom\\
$^2$Institute of Cosmology and Gravitation, University of Portsmouth,
Dennis Sciama Building, Portsmouth, PO1 3FX, United Kingdom}
\date{\today}
\begin{abstract}
We estimate the amount of vorticity generated at second order in
cosmological perturbation theory from the coupling between first order
energy density and non-adiabatic pressure, or entropy, perturbations.
Assuming power law input spectra for the source terms, and working in
a radiation background, we calculate the wave number dependence of the
vorticity power spectrum and 
its amplitude. We show
that the vorticity generated by this mechanism is non-negligible on
small scales, and hence should be taken into consideration in current
and future CMB experiments.
\end{abstract}


\maketitle

\section{Introduction}

Cosmological perturbation theory is a crucial tool for the study of
the early universe. Linear theory (see, e.g.~Refs.~\cite{Lifshitz,
  Bonnor1957, Tomita1967, Bardeen80, KS, MFB, Ruth}) allows us to
model the large scale structure of the universe \cite{Tegmark:2006az}
and calculate the anisotropies of the cosmic microwave background
(CMB) and compare them to the data from experiments, e.g.~{\sc
  Wmap}. However, with the advent of new data sets, such as those from
the full seven year running of {\sc Wmap} \cite{WMAP7} or those
anticipated from the recently launched {\sc Planck} satellite
\cite{planck}, which are superior in both their quality and quantity
to previous ones, we are now in a position to test predictions of
higher order perturbation theory against observational data.
Conversely, perturbation theory beyond linear order is necessary in
order to extract the maximum amount of information from these data
sets, and thus utilise them to their full potential. The second order
theory has been developed by several authors (see
e.g.~Refs.~\cite{Tomita1967, Bruni:1996im,
  Noh:2004bc,Bartolo:2004if, Nakamura, MM2008, MW2008}), 
and there have been studies of the applications of the
theory ranging from calculations of the non-Gaussianity of the
primordial perturbations to investigations of higher order phenomena
produced by couplings between lower order perturbations.

It is well known that at linear order in perturbation theory,
different types of perturbations (classified as scalar, vector and
tensor perturbations, according to their transformation behaviour on
three dimensional hypersurfaces) decouple \cite{KS}.  However, at
second order (see, e.g.~Ref.~\cite{MW2008} and references therein),
and beyond \cite{third}, this is no longer the case, and mode
couplings exist which give rise to novel, interesting features.
One well-studied example of such a higher order effect is that of
gravitational waves sourced by first order density perturbations. This
topic has received a lot of recent attention \cite{Mollerach:2003nq,
  Ananda:2006af, Baumann:2007zm, Assadullahi:2009nf} and the
contribution of the second order tensors to the tensor spectrum has
been studied in some detail.  \\

The production of vorticity at second order in perturbation theory is
another effect that only becomes important at this and higher orders
in a Friedmann-Robertson-Walker background, and is the focus of the
current work.
For the case of a barotropic fluid Ref.~\cite{Lu:2008ju} showed that
vorticity is not produced at any order in perturbation theory.
In Ref.~\cite{vorticity} we showed, dropping this assumption of
barotropicity, that vorticity is indeed generated at second order. We
considered a generic perfect fluid, i.e.~a fluid whose energy momentum
tensor is diagonal (see, e.g.~\Ref{Giovannini:2008zz}). The source
term of the vorticity evolution equation contains terms coupling
the linear energy density and non-adiabatic pressure (entropy)
perturbations. We derived the evolution equation for the second order
vorticity tensor, $\w_{2ij}$, assuming zero first order vorticity and
no anisotropic stress, as
\bea
\label{eq:vorsecondevolution}
\w_{2ij}^\prime -3\H\cs\w_{2ij}
=\frac{2a}{\rhob+\Pb}\left\{3\H V_{1[i}\dPn_{1,j]}
+\frac{\drho_{1,[j}\dPn_{1,i]}}{\rhob+\Pb}\right\}\,,
\eea
where $a(\eta)$ is the scale factor, $\H$ the Hubble parameter, $\cs$
the adiabatic sound speed, $\rhob$ and $\Pb$ are the energy density
and pressure in the background, $V_{1i}$, $\dPn_{1}$, $\delta\rho_1$,
are the covariant velocity, the non-adiabatic pressure, and the energy
density perturbation at first order, respectively, and a prime denotes
differentiation with respect to conformal time.  From this equation,
we can see that even in the absence of anisotropic stress, terms
quadratic in the first order perturbations will act as source terms
for the vorticity tensor at second order. This is very different from the
case at linear order where, in the absence of anisotropic stress, the
vorticity has no source term and will decay with the expansion of the
universe \cite{Bardeen80,KS,LLBook,Lewis:2004kg,Hollenstein:2007kg}.

To keep our results conservative and our calculation as simple as
possible and hence analytically tractable, we assume that the source
term in \eq{eq:vorsecondevolution} is dominated by the second
term. 
Then, choosing the radiation
era as our background in which $\cs=1/3$, the evolution equation
simplifies to
\bea
\label{eq:vorevrad}
\w_{2ij}^\prime -\H\w_{2ij}
=\frac{9a}{8\rhob^2}{\drho_{1,[j}\dPn_{1,i]}}
\,.
\eea
The main goal of this paper is to get a handle on the power spectrum
for the second order vorticity sourced by the coupling between first
order density and non-adiabatic pressure perturbations, using
\eq{eq:vorevrad}. We take as input power spectra for the source terms
simple power laws in time and wavenumber (in Fourier space), and
analytically solve for the evolution of $\w_{2ij}$. \\

The paper is organised as follows. In the next section we introduce
our definitions and present the governing equations, followed by an
analytical solution for the linear energy density perturbation in the
presence of entropy. In Section \ref{sec:vor} we solve for the
evolution of the vorticity. We discuss these results and conclude in
Section \ref{sec:conc}.

In this paper we use conformal time throughout. Latin indices $i, j,
k,$ take the value $1, 2,$ or $3$, and we denote a partial derivative
with a subscript comma. The order of the perturbations is denoted with
a subscript immediately after a perturbed quantity and we work in the
uniform curvature gauge.

\section{Definitions and Equations}

The Friedmann-Robertson-Walker (FRW) metric tensor, up to and
including second order perturbations has, in the uniform curvature
gauge, the line element (see, e.g.~Ref.~\cite{MW2008})
\be
ds^2=a^2(\eta)\left[-(1+2\phi_1+\phi_2)d\eta^2+(2B_{1i}+B_{2i})d\eta dx^i
+\delta_{ij}dx^idx^j\right]\,,
\ee
where we have neglected tensor perturbations, as they will not
 change our results qualitatively, and are considering a flat $(K=0)$
background. Here, $\phi_1$ and $\phi_2$ are the lapse function at
first and second order, respectively, and $B_{1i}$ and $B_{2i}$ denote
the shear in this gauge. We consider the matter content of the
universe to be a perfect fluid, with equation of state
$\Pb=w\rhob$, and spatial three velocity $v_i$.  The covariant
velocity perturbation, $V_i$, is defined as $V_i=B_i+v_i$ \cite{MW2008}.

In the remainder of this section we present the evolution and
constraint equations to first order in cosmological perturbation
theory necessary to derive an analytical solution for the first order
density perturbation in the uniform curvature gauge. Having found the
exact analytical solution we approximate it by a power law valid at
early times.

\subsection{Governing equations}
\label{sec:gov}

The evolution and constraint equations are given by energy-momentum
conservation and the Einstein field equations, respectively. In the
background, these are the familiar continuity and Friedmann equations
of a FRW universe, namely
\begin{align}
\rhob'&=-3\H(\rhob+\Pb) \,,\\
\label{eq:Friedmann}
\H^2&=\frac{8\pi G}{3}a^2\rhob \,.
\end{align}
At first order, the governing evolution and constraint equations
are, in the left- and right-hand columns respectively, 
\begin{align}
&\delta\rho_{1}'
+3\H\left( {\delta\rho_{1}}+{\delta P_{1}}\right)
+\left(\rho_{0}+P_{0}\right) \nabla^2 v_1 
=0\,, &
\label{eq:field1_0i}
&\H\phi_1=-4\pi G
a^2\bar{V}_1\,,&\\
%
&\bar{V}_1'
+4\H\bar{V}_1
+\left(\rho_{0}+P_{0}\right)
\phi_1
+\delta P_1=0\,,&
\label{eq:field1_00}
&3\H^2\phi_1+\H\nabla^2 B_1
=-4\pi Ga^2
\delta\rho_1\,,&
\end{align}
where $\nabla^2\equiv\partial_i\partial^i$ denotes the spatial
Laplacian, and we have introduced the rescaled covariant velocity
perturbation, $\bar{V}_1$, for notational convenience, as
\be
\bar{V}_1\equiv\left(\rho_{0}+P_{0}\right)\left(v_1+B_1\right)\,.
\ee
We refrain from deriving the governing equations \eq{eq:field1_0i} and
\eq{eq:field1_00} here, and instead point the interested reader to,
e.g.~Ref.~\cite{MW2008}, for details. Using the Friedmann equation,
\eq{eq:Friedmann}, the right-hand equation in \eq{eq:field1_0i} can be
rewritten as
\be
\phi_1=-\frac{3}{2}\frac{\H}{\rho_0}\V1\,,
\ee
which can then be used to write the right-hand equation of
\eq{eq:field1_00} as
\be
\nabla^2 B_1=\frac{9}{2}\frac{\H^2}{\rho_0}\V1
-\frac{3}{2}\frac{\H}{\rho_0}\delta\rho_1\,.
\ee
Hence the evolution equations then become
\bea
\label{drho1}
&\delta\rho_1'+\frac{3}{2}\H\left(3+w\right)\delta\rho_1+3\H\delta P_1
- k^2\V1-\frac{9}{2}\H^2\left(1+w\right)\V1=0\,,&\\
\label{dv1}
&\V1'+\frac{1}{2}\H\left(5-3w\right)\V1+\delta P_1=0\,,&
\eea
where we are working in Fourier space, $k$ being a comoving
wavenumber (as usual).
Equations (\ref{drho1}) and (\ref{dv1}) make up a system of coupled,
linear, ordinary differential equations. Given an equation of state
and initial conditions, this system can be solved immediately (for a
given $k$), numerically.  One can also obtain a qualitative solution
by considering a system of two equations such as this one.  However,
if one wants to solve the system quantitatively, and analytically, it
is easier to rewrite the system as a single second order differential
equation, which we do in the following. We solve \eq{drho1} for $\V1$
and get
\be
\label{defV1}
\V1=\frac{1}{X}\left(
\delta\rho_1'+\frac{3}{2}\H\left(3+w\right)\delta\rho_1+3\H\delta P_1
\right)\,,
\ee
where
$
X\equiv k^2+\frac{9}{2}\H^2\left(1+w\right)\,.
$
After some further algebraic manipulations of \eq{defV1}, and using
\eq{dv1}, we arrive at the desired evolution equation
\bea
\label{master}
\delta\rho_1''+\left(7\H-\frac{X'}{X}\right)\delta\rho_1'
&+&\frac{3}{2}\H\left(3+w\right)\left[
\frac{\H'}{\H}+\frac{w'}{(3+w)}-\frac{X'}{X}+\frac{1}{2}\H\left(5-3w\right)
\right]\delta\rho_1\nonumber \\
&+&3\H\delta P_1'
+\left[
3\H\left(\frac{\H'}{\H}-\frac{X'}{X}\right)
+\frac{3}{2}\H^2\left(5-3w\right)+X
\right]\delta P_1=0\,.
\eea
Equation (\ref{master}) is a linear differential equation, of second
order in (conformal) time. It is valid on all scales and for a single
fluid with any (time dependent) equation of state. Furthermore, it
assumes nothing more than a perfect fluid and hence allows for non-zero
non-adiabatic pressure, or entropy,  perturbations.

\subsection{Solutions}

Having derived a general governing equation (\ref{master}) valid in
any epoch, we now restrict our analysis to radiation domination, where
the background equation of state parameter is $w=\frac{1}{3}$ and the
adiabatic sound speed is $\cs=\frac{1}{3}$.  During the radiation era
the scale factor scales as $a\propto\eta$, and hence $\H\propto\eta^{-1}$. We
note that the first order pressure perturbation can be expanded as
(see, e.g.~\Ref{nonad})
\be
\label{dP1split}
\delta P_1\equiv \cs \delta\rho_1 +\dPn_1\,,
\ee
where the non-adiabatic pressure perturbation is defined, in terms of
the energy density, $\rho$, and entropy, $S$, as 
\be 
\dPn_{1}\equiv\left.\frac{\p P}{\p S}\right|_{\rho}\delta S_1\,.
\ee
Then, the general governing equation, \eq{master}, becomes, in radiation 
domination,
\bea
&&\delta\rho_1''+4\H\left(2+\frac{3\H^2}{k^2+6\H^2}\right)\delta\rho_1'
+\left(6\H^2+\frac{1}{3}(k^2+6\H^2)+\frac{72\H^4}{k^2+6\H^2}\right)
\delta\rho_1\nonumber\\
&&
\label{eq:masterrad}
\qquad\qquad\qquad\qquad\qquad\qquad+3\H\dPnad'
+\left[9\H^2+36\frac{\H^4}{k^2+6\H^2}+k^2
\right]\dPn_1
=0\,.
\eea
For the case of zero non-adiabatic pressure perturbations the second
line in \eq{eq:masterrad} vanishes, and the resulting equation can be
solved directly using the Frobenius method, to give
\be
\delta\rho_1({\bm k}, \eta)=C_1({\bm k})\eta^{-5}\Big[\cos\left(\frac{k\eta}{\sqrt{3}}\right)k\eta-2\sin\left(\frac{k\eta}{\sqrt{3}}\right)\sqrt{3}\Big]
+C_2({\bm k})\eta^{-5}\Big[2\cos\left(\frac{k\eta}{\sqrt{3}}\right)\sqrt{3}+\sin\left(\frac{k\eta}{\sqrt{3}}\right)k\eta\Big]\,,
\ee
where $C_1$ and $C_2$ are functions of the wave-vector, ${\bm k}$. For
small $k{\eta}$, the trigonometric functions can be expanded in power
series giving, to leading order, the approximation
\be
\label{eq:drho1}
\drho_1({\bm k},{\eta})\simeq A({\bm k})k{\eta}^{-4}+B({\bm k}){\eta}^{-5}\,,
\ee
for some $A({\bm k})$ and $B({\bm k})$, determined by the initial
conditions.

In order to solve for a non vanishing non-adiabatic pressure, we make
the ansatz that the non-adiabatic pressure grows as the decaying
branch of the density perturbation in \eq{eq:drho1}, i.e.,
\be
\label{eq:ansatz}
\dPn_1({\bm k}, \eta)= D({\bm k}) k^\lambda{\eta}^{-5}\,.
\ee
This assumption is well motivated, since we would expect the
non-adiabatic pressure to decay faster than the energy density. This
gives the solution
\begin{align}
\label{eq:drho1nonad}
\drho_1({\bm k}, \eta)  &={{C_1({\bm k})\,{\eta}^{-5} 
\left[ \cos \left( \frac{k\eta}{\sqrt {3}}
 \right) k\eta-2\,\sin \left( \frac{k\eta}{\sqrt {3}} \right) \sqrt {3} 
\right]
}{}}\nn\\
&\qquad+C_2({\bm k})\,\eta^{-5}{{ \left[ 2\,\cos \left(\frac{k\eta}{\sqrt {3}}
 \right) \sqrt {3}+\sin \left(\frac{k\eta}{\sqrt {3}} \right) k\eta \right] {
}}{}}-3D({\bm k})k^\lambda\eta^{-5}\,.
\end{align}

Anticipating results from the following section, we put the vector dependence
of the $\drho_1$ and $\dP_{\rm{nad}1}$ into Gaussian random variables $\hat{E}({\bm k})$, allowing
us to write, for example, $\drho_1({\bm k}, \eta)=\drho_1({ k}, \eta)\hat{E}({\bm k}),$
where $k=\left|{\bm k}\right|$. Then as a further approximation for the scalar $k$-dependent
source terms, we can expand
Eq.~(\ref{eq:drho1nonad}) to lowest order in $k\eta$, and use the
aforementioned ansatz for the non-adiabatic pressure perturbation,
Eq.~(\ref{eq:ansatz}). Taking the functions $A({k})$ and $D({k})$ to
be power laws in $k$ this gives
\be
\label{input_power}
\drho_1({k}, \eta)
=\bar{A} k^{\beta}\eta^{-4}\,, \hspace{1cm}
\dPn_1({k}, \eta)=\bar{D} {k}^\alpha{\eta}^{-5}\,,
\ee
where $\bar{A}$ and $\bar{D}$ are yet unspecified amplitudes and $\beta$ and
$\alpha$ undetermined powers.

\section{Solving the vorticity evolution equation}
\label{sec:vor}

Having obtained a solution for the first order energy density
perturbation in the previous section, we can now turn towards solving
the evolution equation for the second order vorticity. Note that, to
avoid notational ambiguities, we omit the subscript denoting the order
of the perturbation in the following: the vorticity tensor is a second
order quantity, and the energy density and non-adiabatic pressure
perturbation are first order quantities.

\subsection{The power spectrum of the vorticity}
\label{subsec:powervort}

We now derive the power spectrum for the second order vorticity
tensor. Recall from Eq.~(\ref{eq:vorevrad}) that the vorticity
evolves, during radiation domination, according to
\bea
\label{eq:vorev2}
\w_{ij}^\prime -\H\w_{ij}
=\frac{9a}{8\rhob^2}{\drho_{,[j}\dPn_{,i]}}
\equiv S_{ij}
\,.
\eea
The source term can then be written, in Fourier space, as
\be 
S_{ij}({\bm k}, \eta)=-\frac{f(\eta)}{(2\pi)^{3/2}}
\int d^3\tilde{\bm k}(\tilde{k}_i k_j-\tilde{k}_j k_i)
\dPn(\tilde{\bm k}, \eta)\drho({\bm k}- \tilde{\bm k})\,,
\ee
where we have defined 
\be 
f(\eta)=\frac{9a(\eta)}{16\rhob(\eta)^2}\,.
\ee 
It is 
easier to work with the vorticity vector, obtained by contracting the
vorticity tensor with the totally antisymmetric tensor as 
$\w_i({\bm x})=\varepsilon_{ijk}\w^{jk}({\bm x})$ (see
e.g.~Ref.~\cite{Lu:2008ju}). One can define a source vector in an
analogous way, which can be decomposed in terms of basis vectors as
\begin{align}
S_i({\bm x})=\frac{1}{(2\pi)^{3/2}}\int d^3{\bm k}\,
\S_{A}({\bm k})\,e_i^{A}\,{\rm e}^{i{\bm k}\cdot{\bm x}}
=\frac{1}{(2\pi)^{3/2}}\int d^3{\bm k}
\Big[\S_{1}e_i+\S_2\bar{e}_i+\S_3\hat{k}_i\Big]{\rm e}^{i{\bm k}\cdot{\bm x}}\,.
\end{align}
This enables us to write the evolution equation, Eq.~(\ref{eq:vorev2}), as
\bea
\label{eq:evpol}
\w_{A}'({\bm k}, \eta)-\H\w_{A}({\bm k},\eta)=\S_{A}({\bm k},\eta)\,,
\eea
for each basis state, $A$. We only have to consider the case $A=1$,
and therefore drop the subscripts in the following\footnote{Note that
  we make this assumption without loss of generality: of the two other
  basis states, one is zero from the definition and the other gives
  zero contribution to the power spectrum after the following
  calculations.}.
The power spectrum of the vorticity is defined, in the usual manner, as
\be
\label{eq:PSomega}
\left<{\w}^*({\bm k}_1, \eta){\w}({\bm k}_2, \eta)\right>
=\frac{2\pi}{k^3}\delta({\bm k}_1-{\bm k}_2){\mathcal{P}}_\w(k, \eta)\,,
\ee
where the asterisk denotes the complex conjugate. Since
Eq.~(\ref{eq:evpol}) is a simple first order differential equation in
conformal time, it can be integrated directly, which then enables us
to write the correlator for the vorticity as
\be
\label{eq:PSomegawithS}
\left<{\w}^*({\bm k}_1, \eta){\w}({\bm k}_2, \eta)\right>
=\eta^2\int_{\eta_0}^\eta \tilde{\eta}_1^{-1}
\int_{\eta_0}^\eta \tilde{\eta}_2^{-1} 
\left<{\S}^*({\bm k}_1, \tilde{\eta}_1){\S}({\bm k}_2, \tilde{\eta}_2)\right> 
d\tilde{\eta}_1d\tilde{\eta}_2\,,
\ee
where $\eta_0$ is a constant denoting an initial time. Thus, we can
relate the power spectrum, ${\mathcal{P}}_\w(k, \eta)$, to the
correlator of the source terms by equating Eq.~(\ref{eq:PSomega}) and
Eq.~(\ref{eq:PSomegawithS}). Then, assuming that the fluctuations
$\delta\rho$ and $\delta P_{\rm{nad}}$ are Gaussian, we can put the
directional dependence into Gaussian random variables $\hat{E}({\bm
  k})$, for example, by writing $\delta\rho({\bm k},
\eta)=\delta\rho(k,\eta)\hat{E}({\bm k})$. These Gaussian random
variables obey the relationship 
$
\big<\hat{E}({\bm k}_1)\hat{E}({\bm k}_2)\big>
=\delta^3({\bm k}_1+{\bm k}_2).
$
Doing so enables us to evaluate the correlator of the source terms, 
$\big<{\S}^*({\bm k_1}, \eta){\S}({\bm k_2}, \eta)\big>$,
and obtain, after some algebra, the
expression for the power spectrum of the vorticity
\begin{align}
\mathcal{P}_\w(k,\eta)&=\frac{k^5\eta^2}{4\pi^4}\int f(\tilde{\eta}_1)
f(\tilde{\eta}_2)\tilde{\eta}_1^{-1}\tilde{\eta}_2^{-1}
d\tilde{\eta}_1 d\tilde{\eta}_2
\int d^3\tilde{k}(\bar{e}_i \tilde{k}^i)^2
\dPn(\tilde{k},\tilde{\eta}_1)\drho(|{\bm k}+\tilde{\bm k}|,\tilde{\eta}_1)\nn\\
&\qquad\qquad\qquad\times\Big[\dPn(|{\bm k}+\tilde{\bm k}|, \tilde{\eta}_2)\drho(\tilde{k},\tilde{\eta}_2)
-\dPn(\tilde{k},\tilde{\eta}_2)\drho(|{\bm k}+\tilde{\bm k}|, \tilde{\eta}_2)\Big]\,.
\end{align}
Substituting the approximations (\ref{input_power}) into
the above  gives
\begin{align}
\label{eq:PSetaint}
\mathcal{P}_\w(k,\eta)&=\frac{81}{256}\frac{k^5\eta^2}{4\pi^4
}(AD)^2
\left[\ln\left({\frac{\eta}{\eta_0}}\right)\right]^2
\int d^3\tilde{k}(\bar{e}_i \tilde{k}^i)^2\tilde{k}^\alpha|{\bm k}+\tilde{\bm k}|^\beta
\Big(|{\bm k}+\tilde{\bm k}|^\alpha\tilde{k}^\beta-\tilde{k}^\alpha|{\bm k}+\tilde{\bm k}|^\beta\Big)
\end{align}
where we have performed the temporal integral by noting that
as mentioned above,
$a\propto\eta$ during radiation domination, and thus $\rhob\propto\eta^{-4}$.
To perform the $k$-space integral, we first move to spherical coordinates
oriented with the axis in the direction of ${\bm k}$.  Then, denoting
the angle between ${\bm k}$ and ${\tilde{\bm k}}$ as $\theta$, the
integral can be transformed as
\be
\int d^3\tilde{k} \to 2\pi\int_0^{k_{\rm c}}\tilde{k}^2d\tilde{k}\int_0^\pi\sin\theta d\theta\,,
\ee
where the prefactor comes from the fact that the integrand has no
dependence on the azimuthal angle, and $k_{\rm c}$ denotes a cut-off on small scales. 
Noting that, in this coordinate
system $\tilde{k}_i\bar{e}^i=\tilde{k}\sin\theta$
the integral in Eq.~(\ref{eq:PSetaint}) becomes
\begin{align}
\label{eq:int}
I(k)&=2\pi\int_0^{k_{\rm c}}\int_0^\pi\tilde{k}^{4+\alpha}\sin\theta \, d\theta\,
 d\tilde{k}\,\sin^2\theta
k^\beta\Big[1+(\tilde{k}/k)^2+2(\tilde{k}/k)\cos\theta\Big]^{\beta/2}\nn\\
&\qquad\qquad\qquad\times\Big(k^\alpha(1+(\tilde{k}/k)^2+2(\tilde{k}/k)\cos\theta)^{\alpha/2}\tilde{k}^\beta-
\tilde{k}^\alpha k^\beta (1+(\tilde{k}/k)^2+2(\tilde{k}/k)\cos\theta)^{\beta/2}\Big)
\end{align}
Finally, in order to solve this integral we change variables again to
dimensionless $u$ and $v$ defined as \cite{Ananda:2006af} (or similarly \cite{Brown:2010ms})
\begin{align}
v=\frac{\tilde{k}}{k}\,, \, \,
u=\sqrt{1+(\tilde{k}/{k})^2+2(\tilde{k}/{k})\cos\theta}=\sqrt{1+v^2+2v\cos\theta}\,,
\end{align}
for which the integral (\ref{eq:int}) becomes
\begin{align}
\label{integralIk}
I(k)
&=k^{2(\alpha+\beta)+5}\int_0^{{k_{\rm c}}/{k}}\int^{v+1}_{|v-1|}u\, du\, v^3\, dv\, u^\beta v^\alpha
\Big(1-\frac{1}{4v^2}(u^2-v^2-1)^2\Big)
\Big[u^\alpha v^\beta -v^\alpha u^\beta \Big]\,.
\end{align}
%

\subsection{Evaluating the vorticity power spectrum}
\label{subsec:evalpowervort}

In order to perform the integral \eq{integralIk} derived above, we need
to specify the exponents for the power spectra of the energy density
and the non-adiabatic pressure $\alpha$ and $\beta$. The energy
density perturbation can be related to the curvature perturbation on
uniform density hypersurfaces, $\zeta$, during radiation domination
through \cite{MW2008}
\be
\delta\rho=-\frac{\rho_0'}{\H}\zeta=4\rhob\zeta\,,
\ee
and hence the initial power spectra can be related as
$\left<\delta\rho_{\rm ini}\delta\rho_{\rm ini}\right>
=16\rho_{0\gamma\rm ini}^2 \left<\zeta_{\rm ini}\zeta_{\rm
ini}\right>$, and we get the power spectrum of the initial
density perturbation 
\be
\delta\rho_{\rm ini}\propto\zeta_{\rm ini}
\propto\left(\frac{k}{k_0}\right)^{\frac{1}{2}(n_{\rm s}-1)}\,,
\ee
where $k_0$ is the {\sc{Wmap}} pivot scale and $n_{\rm s}$ the
spectral index of the primordial curvature perturbation \cite{WMAP7}.
This allows us to relate our ansatz for the density perturbation,
\eq{input_power}, to the {\sc{Wmap}}-data
which gives
\be 
\delta\rho=\delta\rho_{\rm ini}\,\left(\frac{k}{k_0}\right)\left(\frac{\eta}{\eta_0}\right)^{-4}
=A_{\rm ini}\rho_{0{\rm ini}}
\left(\frac{k}{k_0}\right)^{\frac{1}{2}(n_{\rm s}+1)}
\left(\frac{\eta}{\eta_0}\right)^{-4}\,.
\ee
From this, we can read off that $\beta=\frac{1}{2}(n_{\rm s}+1)\simeq
1$ and the amplitude $A=A_{\rm ini}\rho_{0{\rm ini}}$.
 We have some freedom in
choosing $\alpha$, however would expect the non-adiabatic pressure to
have a blue spectrum, though the calculation demands
$\alpha\neq\beta$.
Using the notation of Ref.~\cite{WMAP7} we get
\be 
A^2=\rho_{0  \rm ini}^2
{\mathcal{P}}_{\mathcal{R}}(k_0)^2=
\rho_{0  \rm ini}^2k_0^{-6}
\Delta_{\mathcal{R}}^4\,, \hskip 1cm 
D^2=\rho_{0 \rm ini}^2
{\mathcal{P}}_{\mathcal{S}}(k_0)^2
= \rho_{0 \rm ini}^2 k_0^{-6}
\Delta_{\mathcal{S}}^4
\ee 
where we also have the ratio
\be 
\frac{\Delta_{\mathcal{S}}^2}{\Delta_{\mathcal{R}}^2}
=\frac{\alpha(k_0)}{1-\alpha(k_0)}\,,
\ee
and therefore, 
\be 
(AD)^2
=\frac{\alpha(k_0)}{1-\alpha(k_0)}\Delta_{\mathcal{R}}^8\rho_{0 \rm ini}^4k_0^{-12}\,,
\ee
where we can substitute in numerical values for $\Delta_{\mathcal{R}}^2$ and
$\alpha(k_0)$ from Ref.~\cite{WMAP7} later on.

Then, making the choice $\alpha=2$, the input
power spectra are 
\be
\label{input_power_norm}
\drho({k}, \eta)
=A \left(\frac{k}{k_0}\right)\left(\frac{\eta}{\eta_0}\right)^{-4}\,, \hspace{1cm}
\dPn({k}, \eta)= D\left(\frac{k}{k_0}\right)^2\left(\frac{\eta}{\eta_0}\right)^{-5}\,,
\ee
for which the integral \eq{integralIk} becomes
\begin{align}
I(k)
&={k^{11}}
\int_0^{{k_{\rm c}}/{k}}\int^{v+1}_{|v-1|}u^2\, du\, v^5\, dv\, 
\Big(1-\frac{1}{4v^2}(u^2-v^2-1)^2\Big)
\Big[u^2 v -v^2 u\Big]\,.
\end{align}
We can then integrate this analytically, taking care when it comes to evaluating
the limits, to give
\be 
\label{integrated}
I(k)=\frac{16}{135}k_{\rm c}^9k^2+\frac{12}{245}k_{\rm c}^7k^4-\frac{4}{1575}k_{\rm c}^5k^6\,,
\ee 
which clearly depends upon the small scale cut-off, as expected. For illustrative 
purposes, let's choose $k_{\rm c}=10$ and plot the solution $I(k)$ in Figure 1.
The left hand figure shows that the amplitude of the integral grows as the wavenumber
increases.  The right hand figure shows a turn
around and a decrease in power at some wavenumber (in fact, for a non-specific cutoff,
this point is at $3.7375k_{\rm c}$). However, we note that this 
value is greater than our cutoff, and therefore unlikely to be physical.
\begin{figure}[ht]
\subfigure[Small range of $k<k_{\rm c}$.]{
\includegraphics[width=0.48\textwidth]{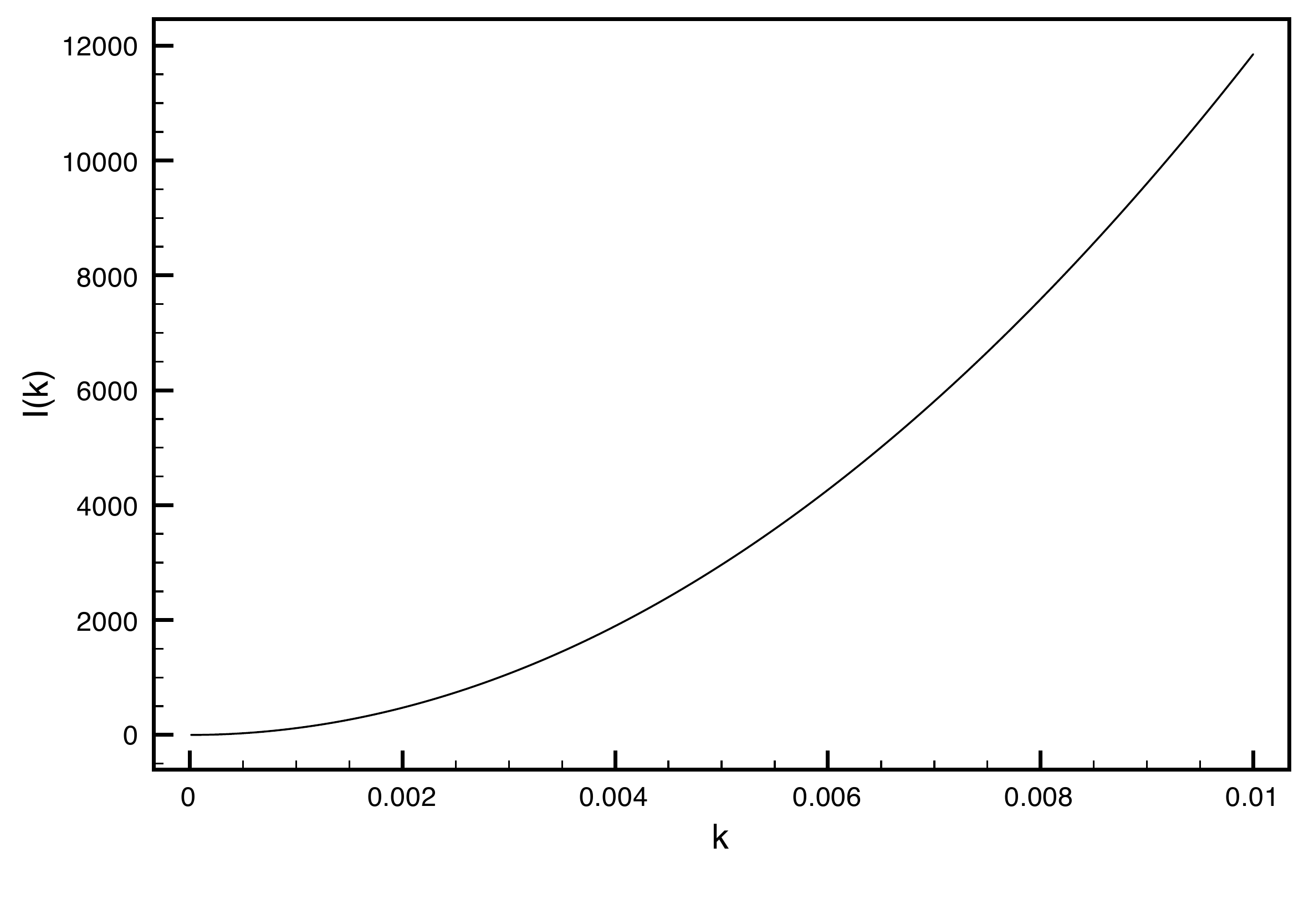}
}
\subfigure[ Wide range of $k$ including $k>k_{\rm c}$.]{
\includegraphics[width=0.48\textwidth]{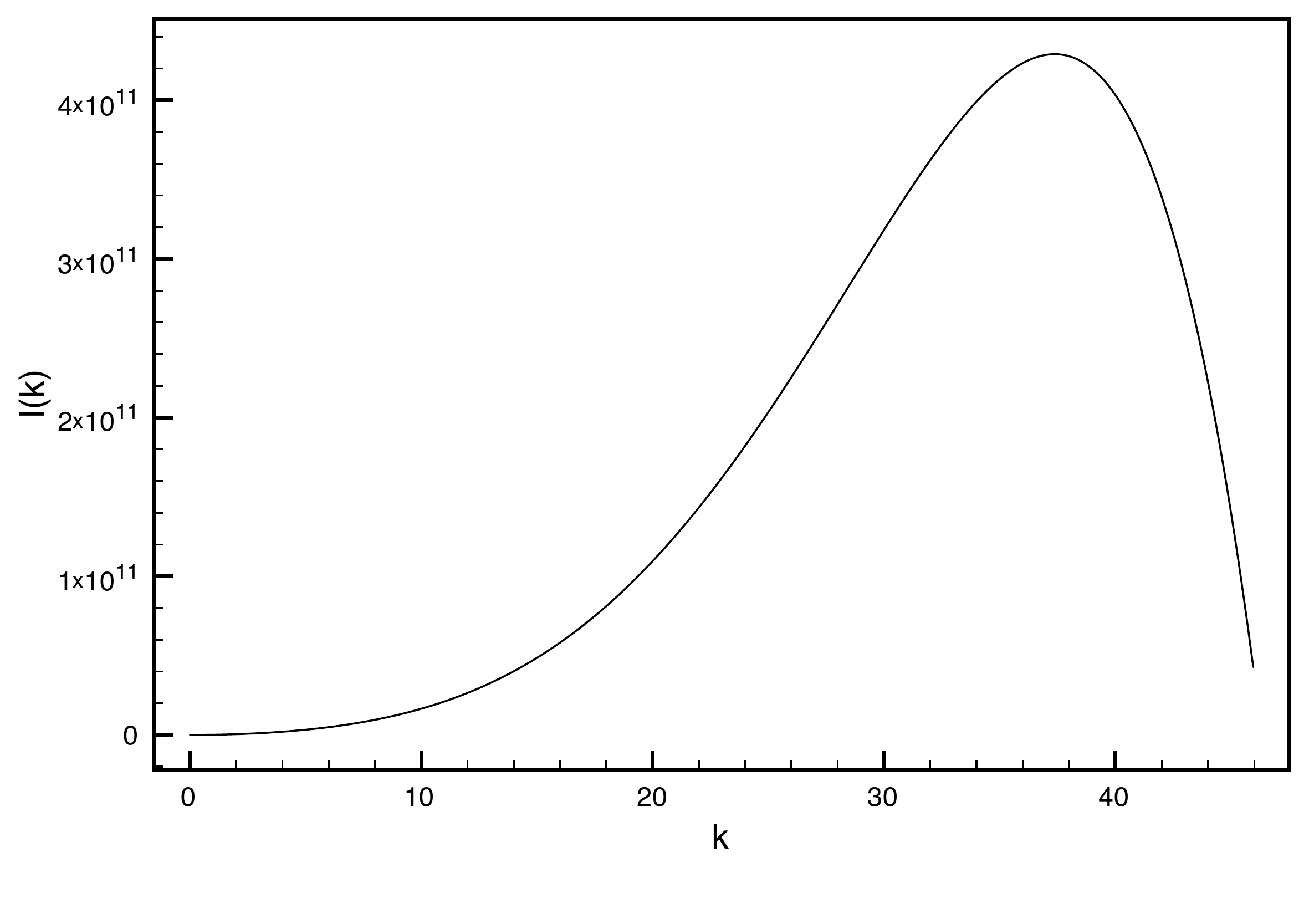}
}
\caption{Plots of $I(k)$ for the illustrative choice of $k_{\rm c}=10$.}
\end{figure}

Then, using the above, and noting that we must use the input to the temporal integrals as
\be 
a\propto\left(\frac{\eta}{\eta_0}\right)\,,
\hskip 1cm
\rho_0=\rho_{0 \rm ini}\left(\frac{\eta}{\eta_0}\right)^{-4}\,,
\ee
we obtain the power spectrum for the vorticity, for general $k_{\rm c}$ as 
\begin{align}
{\mathcal{P}}_\w(k, \eta)&=
\frac{81}{256}k_0^5\frac{\eta^2}{4 \pi^4}\frac{1}{\rho_{0 \rm ini}^4}\ln^2\left(\frac{\eta}{\eta_0}\right)
\Bigg(\frac{\alpha(k_0)}{1-\alpha(k_0)}\Bigg)^2k_0^{-12}\Delta_{\mathcal{R}}^8\rho_{0 \rm ini}^4k_{\rm c}^5
\Bigg[\frac{16}{135}\frac{k_{\rm c}^4}{k_0^4}\left(\frac{k}{k_0}\right)^7
+\frac{12}{245}\frac{k_{\rm c}^2}{k_0^2}\left(\frac{k}{k_0}\right)^9
-\frac{4}{1575}\left(\frac{k}{k_0}\right)^{11}\Bigg]\,,
\end{align}
and on substituting in values from Ref.~\cite{WMAP7}, taking a conservative estimate for $\alpha(k_0)$, being
10 per cent of the upper bound
\begin{align}
{\mathcal{P}}_\w(k, \eta)
&=\eta^2\ln^2\left(\frac{\eta}{\eta_0}\right)\Bigg[0.87 \times 10^{-12} 
k_{\rm c}^9\left(\frac{k}{k_0}\right)^7
{\rm Mpc^{11}}
+3.73\times 10^{-18}k_{\rm c}^7\left(\frac{k}{k_0}\right)^9{\rm Mpc^{9}}\nn\\
&\qquad\qquad\qquad\qquad\qquad
-7.71\times 10^{-25}k_{\rm c}^5 \left(\frac{k}{k_0}\right)^{11}{\rm Mpc^{7}}\Bigg]
\,,
\end{align}
and for our above choice of $k_{\rm c}=10 {\rm Mpc}^{-1}$,
\be 
{\mathcal{P}}_\w(k, \eta)=\eta^2\ln^2\left(\frac{\eta}{\eta_0}\right)\Bigg[0.87\times 10^{-2}
\left(\frac{k}{k_0}\right)^7
+3.73\times 10^{-11}\left(\frac{k}{k_0}\right)^9
-7.71\times 10^{-20}\left(\frac{k}{k_0}\right)^{11}\Bigg]{\rm Mpc^{2}}\,.
\ee

\section{Discussion and conclusion}
\label{sec:conc}

In this paper we have obtained the first realistic calculation of the
amount of vorticity generated at second order in cosmological
perturbation theory by allowing for a non-adiabatic pressure
perturbation, as would be the case in a multi-field or multi-fluid
system.
By solving the governing equations at first order, we have been able
to obtain an approximate input power spectrum for the energy density
perturbation. Then, making the ansatz that the non-adiabatic pressure
perturbation has a bluer spectrum than that of the energy density in
order to keep the non-adiabatic pressure sub-dominant on all scales, we obtain
an analytical result for the vorticity.
Our results show that the vorticity power spectrum has a non-negligible magnitude which
 depends on the cut off, $k_{\rm c}$ and the chosen parameters. 
As this is a second order effect the magnitude is somewhat surprising.
 We have also shown that the result has 
a dependence on the wavenumber to the power of at least seven for the choice
$\alpha=2$.
Therefore the amplification due to the large power of
$k$ is huge, rendering the vorticity not only possibly observable
but also important for the general understanding of the physical
processes taking place in the early universe.

As discussed above, we have only approximated the input power spectra. 
This approximation necessarily only holds for some time
and some range of wavenumbers.
The next step will be to solve the system of equations numerically,
both the vorticity evolution equations and the input power
spectra. This can be done for various settings, e.g.~by calculating
$\dPn$ from multi-field scalar field model \cite{Gordon:2000hv} at or
during the end of inflation, or for multi-fluid system during later
epochs, evolving $\delta\rho$ and using \eq{master}.\\

One prospect for observing early universe vorticity is in the
B-mode polarisation of the CMB. Both vector and tensor perturbations
produce B-mode polarisation, but at linear order such vector modes decay with
the expansion of the universe. However, vector modes produced
by gradients in energy density and entropy perturbations, such as those 
discussed in this work, will source B-mode polarisation at second order.
Furthermore, recently it has been noted that vector perturbations in fact 
generate a stronger B-mode polarisation that tensor modes with the 
same amplitude \cite{GarciaBellido:2010if}. Therefore, it is feasible 
for vorticity to be observed by future surveys. Finally, an important consequence 
of vorticity is the generation of magnetic fields, and so early universe 
vorticity could play a crucial role in determining the origin of 
primordial magnetic fields. We will investigate this in a future publication
\cite{inprep}.

\acknowledgments

The authors are grateful to Laila Alabidi,
Iain Brown, Martin Bucher, Ian Huston and David Seery
for useful discussions and comments.
AJC also thanks
Scott Dodelson and Albert Stebbins for interesting discussions and
hospitality in the Astrophysics Theory Group at Fermilab. AJC and KAM
are grateful to the organisers and participants of the GC2010 long
term workshop YITP-T-10-01 for informative discussions and for
providing a stimulating working environment.  AJC is supported by the
Science and Technology Facilities Council (STFC). KAM is supported, in
part, by STFC under Grant ST/G002150/1.  We used the computer algebra
package {\sc{Cadabra}} \cite{Cadabra} to obtain some of the equations
in Section \ref{sec:gov}.



\end{document}